\documentclass[11pt]{article} %%Font size and document presets
\usepackage{enumitem}%%Enables control over enumerate and itemize environments
\setenumerate{itemsep=1pt, topsep=1pt}
\usepackage{setspace} %%Enables \doublespacing command for double linespacing
\expandafter\def\expandafter\quote\expandafter{\quote\onehalfspacing} %%Makes 1.5 quote spacing
\usepackage{graphicx}

\usepackage{fancyhdr} %%Permits \pagestyle{fancy}
\usepackage{titlesec} %%Header style
\titlespacing*{\subsection}{\parindent}{.25in}{\wordsep}% Reduces spacing after headings
\usepackage{scrextend} %%Allows for changes to footnotes
\deffootnote[1em]{0in}{1em}{\textsuperscript{\thefootnotemark \ }} %%Footnote style
\usepackage{amssymb, amsmath, mathrsfs, amsthm, braket} %%Math packages
\usepackage{stmaryrd} %%Use \llbracket and \rrbracket for double brackets

\usepackage[round]{natbib} %%Or change 'round' to 'square' for square backers
\usepackage{etoolbox} %%For \citepos
\usepackage{xstring} %%For \citepos

\makeatletter %definition of \citepos
\DeclareRobustCommand\citepos % define \citepos
  {\begingroup
   \let\NAT@nmfmt\NAT@posfmt% same as for citet except with a different name format
   \NAT@swafalse\let\NAT@ctype\z@\NAT@partrue
   \@ifstar{\NAT@fulltrue\NAT@citetp}{\NAT@fullfalse\NAT@citetp}}
   
\let\NAT@orig@nmfmt\NAT@nmfmt %makes adapt to last names ending with an 's'.
\def\NAT@posfmt#1{%
  \StrRemoveBraces{#1}[\NAT@temp]%
  \IfEndWith{\NAT@temp}{s}
    {\NAT@orig@nmfmt{#1'}}
    {\NAT@orig@nmfmt{#1's}}}
\makeatother

\usepackage{csquotes}
\MakeOuterQuote{"}

\begin{document}
\title{\Large Does Quantum Gravity Happen at the Planck Scale?}
\author{Caspar Jacobs}
\date{\today}
\maketitle
\thispagestyle{empty}

\begin{abstract}
\noindent The claim that at the so-called Planck scale our current physics breaks down and a new theory of quantum gravity is required is ubiquitous, but the evidence is shakier than the confidence of those assertions warrants. In this paper, I survey five arguments in favour of this claim - based on dimensional analysis, quantum black holes, generalised uncertainty principles, the nonrenormalisability of quantum gravity, and theories beyond the standard model - but find that none of them succeeds. The argument from nonrenormalisability is the most convincing, yet it requires the unwarranted assumption that the same constant of action occurs in every quantum field theory. Therefore, our theories don't (yet) predict that quantum gravity happens at the Planck scale.
\end{abstract}

%\doublespacing

%%%%%%%%%%%%%%%%%%%%%%%% NOTES %%%%%%%%%%%%%%%%%%%%%%%%%

%%%%%%%%%%%%%%%%%%%%%%% DOCUMENT %%%%%%%%%%%%%%%%%%%%%%%

\section{Introduction}

The claim that at the so-called Planck scale our current physics breaks down and a new theory of quantum gravity becomes necessary is ubiquitous. In a reference article, \citet{Weinstein2024} write:

\begin{quote}
It is almost Gospel that quantum gravity is what happens when you reach the Planck scale. The standard refrain is that `something peculiar' happens to our concepts of space, time, and causality at such such scales requiring radical revisions that must be described by the quantum theory of gravity.
\end{quote}
However, they also admit that "the arguments underlying this orthodoxy have not been rigorously examined." The aim of this paper is to carry out such an examination. 

The result is that the evidence for quantum gravity at the Planck scale relies on heuristics more than on proof. I therefore concur with \citet{Meschini2007}, who concludes that we should see "the relevance of Planck-scale physics" as a "humble belief, and not as an established fact," which "rests on several uncritical assumptions." I improve on Meschini's analysis in two ways: the more precise characterisation of the supposed relevance of the Planck scale (\S3), and the comprehensive range of arguments I discuss---in particular considerations based on effective field theories.

I will discuss five different arguments for the claim that quantum gravity happens at Planck scale (\S4). These are based on: dimensional analysis, quantum black holes, generalised uncertainty principles, the nonrenormalisability of quantum gravity, and theories beyond the standard model. The argument from nonrenormalisability is the most convincing, yet still requires a non-trivial assumption.

In the conclusion (\S5), I will return to the heuristic role of such assumptions. Do we have reason to believe that they are truth-conductive, or is Planck-scale physics a castle in the air? It is my humble belief that claims about the Planck scale deserve more scepticism than they have yet met with. Given the incredible smallness of the Planck scale, this is clearly of interest in itself: quantum gravity may turn out to become relevant at much larger scales, or at even smaller ones. But the question is not just idle curiosity, since assumptions about the Planck scale may implicitly steer physics practice. On the practical side, physicists hope to `probe the Planck scale' in order to gather data for a future theory of quantum gravity \citep{Pikovski2012, Das2021}. If the Planck scale is not the relevant scale, then some of those efforts are in vain. On the theoretical side, the Planck length is used as a parameter in beyond-the-standard-model theories such as string theory and loop quantum gravity (see \S4.5).  If the Planck length is not the physically relevant length-scale, then predictions of those theories based on a Planck-scale parameter are faulty. Currently, such theories don't have any testable predictions, so the issue has no practical relevance. But that is only a matter of the technology available. The question of whether quantum gravity happens at the Planck scale may therefore well have direct effects on the empirical adequacy of future physics.

\section{Planck Units}

In 1899, Max Planck proposed a system of units based on the natural constants $G$, $c$ and $\hbar$ \citep[\S26]{Planck1899}. The system is based upon the insight that there are unique combinations of those constants with dimensions of length, mass and time:

\begin{align}
\ell_P &= \sqrt{\frac{\hbar G}{c^3}}\\
t_P &= \sqrt{\frac{\hbar G}{c^5}}	\\
m_P &= \sqrt{\frac{\hbar c}{G}}
\end{align}
The Planck unit of length, $\ell_P$, has a value of approximately $1.62 \times 10^{-35}$ m in SI units. In Planck units, $\ell_P$, $t_P$ and $m_P$ are equal to $1$, or equivalently, $G = c = \hbar = 1$. For example, 1 metre is  $6.19 \times 10^{34}$ Planck lengths.

Planck called his units `natural' because don't require any arbitrary artefacts, such as the infamous standard metre in Paris. They are "independent of special bodies or substances, necessarily retaining their meaning for all times and for all civilizations, including extraterrestrial and non-human ones" \citep[174]{Planck1914}. \citet{Grozier2020} traces the widespread adoption of natural units in particle physics to the fact that they simplify equations as $c$ and $\hbar$ disappear from them.

These considerations are unrelated to quantum gravity. Indeed, Planck proposed his units in 1899, well before the discovery of either quantum field theory or general relativity! Nevertheless, it was quickly suggested that they were relevant to future physics. Here is Arthur Eddington, in 1918:
 
\begin{quote}
From the combination of the fundamental constants, $G$, $c$, and $h$ it is possible to form a new fundamental unit of length $L_{min}$ [...]. It seems inevitable that this length must play some role in any complete interpretation of gravitation. \citep[36]{Eddington1918}
\end{quote}
The notion of Planck-scale physics was then further developed by Matvei Bronstein, Oskar Klein and John Wheeler \citep{Gorelik1992}. In more recent times, it has even been suggested that the Planck length is nature's fundamental length scale \citep{Hossenfelder2013}.

How could Planck's values transform from a natural choice of units into a scale set by nature? Or, more concretely: why should we believe that quantum gravity happens at the Planck scale? 

\section{The Claim}

Before I start, I make two clarifications. Firstly: what does it mean for quantum gravity to `happen' at the Planck scale? The slightly less imprecise answer is that `new physics' appears at this scale, that is, physical phenomena not predicted by our present theories. This is an instance of what \citet{Crowther2023} calls `empirical inconsistency'. But absent any means to empirically access this scale, it is hard to assess this. 

The other two types of consistency discussed by Crowther are more relevant: \emph{external} inconsistency concerns the incompatability between different theories, while an \emph{internal} inconsistency is a tension within one theory. Crowther identifies external inconsistency as the most important constraint on quantum gravity, in particular the incompatability between quantum mechanics and general relativity. But as \citet{Wallace2022b} argues, we in fact possess a fine-for-all-practical-purposes theory that combines the two: low-energy quantum gravity. The reason this is not a satisfactory endpoint is that the theory is supposed to become \emph{internally} inconsistent at some scale $\ell_{QG}$. In particular, empirical predictions obtained by perturbative expansions yield untameable infinities at very small scales. The theory is inapplicable at those scales and a new theory of high-energy quantum gravity becomes necessary.  In what follows, I will take this as the relevant type of inconsistency. (Crowther considers this a case of empirical inconsistency, since we can be confident that predictions of infinity are not empirically accurate. To me, the absence of any useful quantitative predictions seems more severe than merely incorrect predictions, but not much depends on this point.)

Secondly, what does it mean to say that quantum gravity happens \emph{at} the Planck scale? There are three interpretations: either the Planck scale is a \emph{lower bound}, an \emph{upper bound} or a \emph{boundary}. If it is a lower bound, then one will definitely encounter quantum gravity by the time one reaches the Planck scale, but it may happen well before that scale: $\ell_{QG} \geq \ell_P$. Conversely, if it is an upper bound then one will not encounter quantum gravity before one reaches the Planck scale but it may only happen at much smaller scales: $\ell_{QG} \leq \ell_P$. Finally, if the Planck scale is a boundary then $\ell_P$ is the more-or-less exact scale at which our current physics breaks down: $\ell_{QG} \approx \ell_{P}$. Although the latter is often intended, we will see that many cases can at most establish one of the former, weaker claims---in particular that the Planck scale is merely a lower bound.

Therefore, by `quantum gravity happens at the Planck scale', I shall mean the conjunction of:

\begin{description}[style=multiline, leftmargin=3.5cm, labelwidth=3cm]
	\item[(Inconsistency)] Physics becomes internally inconsistent at some small length scale; and
	\item[(Boundary)] The scale at which that happens is approximately equal to the Planck scale.	
\end{description}
 In the remainder I will refer to the conjunction of these claims as \emph{Planck}.
 
 This characterisation is consistent with the way many physicists talk about the Planck scale: \citet[1180]{Landau1954} (translated by \citet{Gorelik1992}) state that the Planck length "is the limit of the region outside of which quantum electrodynamics cannot be considered as a self-consistent theory because of the necessity of taking into account gravitational interactions"; \citet[1287]{Rovelli2007} writes that "these theories become meaningless in the regimes where relativistic quantum gravitational effects are expected to become relevant"; and \citet[798]{Peskin1995}, somewhat carefully, claim that "it seems equally probable that quantum field theory will actually break down at [the Planck scale]."

That said, some physicists only commit to claims that are weaker than \emph{Planck}. \citet[71]{Weinberg1981}, for example, states that "[i]n order to avoid an inconsistency between quantum mechanics and general relativity, some new features must enter physics at some energy at or below [the Planck mass]." He thus endorses (Inconsistency), but rejects (Boundary) in favour of the Planck scale as a lower bound (note that a lower Planck mass corresponds to a higher Planck length). \citet{Schwartz2013} even declares that "[t]here is nothing inconsistent about general relativity and quantum mechanics."  It is possible that our theories display pathological features at the Planck scale short of inconsistency, which nonetheless motivate the search for new physics at the Planck scale. I will return to this issue in the conclusion. Still, \emph{Planck}---and (Inconsistency) in particular---seems relatively well-entreched. I will now show that its assertion is far from justified.

\section{Arguments and Refutations}

In this section, I critically discuss a number of arguments for \emph{Planck}, arguing that they fail to establish Inconsistency, Boundary, or both.

\subsection{Dimensional analysis}

The most widespread argument for \emph{Planck} is based on dimensional analysis: \citet[37]{Butterfield2001} refer to "a simple dimensional argument that quantum gravity has a natural length scale---the Planck length"; see also \citet{Baez2001}, \citet{Rovelli2007} and \citet{Rickles2008}. But as \citet{Weinstein2024} point out, "the details of these dimensional arguments, and the role of the Planck scale are calling out for a closer analysis." 

Recall that dimensional analysis is a technique that enables one to derive the functional dependence of one quantity on a set of others based on their dimensions alone. For example, if we know that the period $P$ of a pendulum depends at most on the length of the string, $l$, the mass of the weight, $m$, and the gravitational acceleration, $g$, then dimensional analysis reveals that the only product of powers of those quantities with dimensions of time is $\sqrt{l/g}$---to which $P$ must therefore be proportional.\footnote{Dimensional analysis is a topic of philosophical interest in itself; for recent discussions, see \citet{Lange2009}, \citeauthor{Jallohforthcoming} (forthcoming), \citet{Jacobs2024}, \citet{Skow2017}.}

Dimensional analysis is used to justify \emph{Planck} as follows: suppose one can derive a length-dimensioned quantity $\ell_{QG}$ that sets the scale of quantum gravity. If $\ell_{QG}$ depends on $G$, $c$ and $\hbar$ only, then by dimensional analysis it is proportional to the Planck length. It is often supposed that $\ell_{QG}$ is derivable from a future theory of quantum gravity. \citet[179]{Baez2001}, for example, states that it "is reasonable to suspect that any theory reconciling general relativity and quantum theory will involve all three constants $c$, $G$, and $\hbar$."   But we know far too little about future physics to be confident that exactly those constants appear in it: "it may be required that one or more presently unknown natural constants be discovered and taken into account in order to picture the new physics correctly; or yet more unexpectedly, it may also turn out that quantum gravity has nothing to do with one or more of the above-mentioned constants" \citep[278]{Meschini2007}.\footnote{\citet{Weinstein2024} point out that future theories should recover these constants in the low-energy limit, but that does not mean they directly involve them.} After all, theory change is often associated with the introduction of a novel constant, as was the case for quantum mechanics ($\hbar$) and special relativity ($c$) \citep{Levy-Leblond2019}.

In light of the clarifications in the previous section, however, it is more helpful to think of $\ell_{QG}$ as a prediction of our current theories, namely the scale at which they become internally inconsistent. This obviates the need for speculation, since the constants appear in present theories of low-energy quantum gravity. While Meschini's criticisms are correct, then, they are aimed at a different construal of what it means for quantum gravity to happen at the Planck scale.

With this in mind, here is a reconstruction of the dimensional argument:

\begin{enumerate}
	\item The only length-dimensioned quantity definable from just $G$, $c$ and $\hbar$ is of the order of the Planck length;
	\item The prediction of $\ell_{QG}$ involves $G$, $c$ and $\hbar$---and no other variables;
	\item Therefore, $\ell_{QG}$ is of the order of the Planck length.
\end{enumerate}
But the argument is unsound because neither premiss is true. The first premiss is false since one can multiply the Planck length by any dimensionless constant: the result still depends only on $G$, $c$ and $\hbar$. Baez acknowledges this loophole: "The `unimportant numerical factor' I mentioned above might actually be very large, or very small" (180). The assumption that dimensionless parameters are of order 1 is sometimes known as `naturalness'. This assumption is highly controversial: \citet{Dirac1937} defended it for aesthetic reasons, but \citet{Hossenfelder2017} dismisses it as `numerology' and \citet[13]{Wilson2005} states that it "makes no sense when one becomes familiar with the history of physics". I lack the space to adjudicate this debate here.\footnote{But see \citet{Williams2019a} for a defence of a more sophisticated notion of naturalness.}  Granted, in most concrete applications such numerical factors \emph{are} close to unity; for example in the black hole example discussed below. But not always: in string theory, it is possible that the dimensionless coupling constant that relates the Planck length to the string length is much larger than 1 (see \S4.5). That would invalidate Boundary. This premiss thus relies on a controversial heuristic that may not apply here.

The second premiss is also problematic. While it does not rely on speculation about future physics, \emph{pace} Meschini, it ignores the fact that even our current theories involve many more constants than just $G$, $c$ and $\hbar$: there are the elementary charge, $e$, the mass of the electron, $m_e$, and the proton, $m_p$, and the cosmological constant $\Lambda$, to name a few. From these constants one can construct different sets of units. For example, \citet{Wilczek2005} shows that one can define a complete system of `natural' units from $G$, $c$ and $e$; or from $\hbar$, $e$ and $m_e$; or $\hbar$, $c$ and $m_p$. In the latter system, for example, the  length unit is $\hbar/m_c = 2.1 \times 10^{-14}$ cm. These `strong units', Wilczek suggests, are "more fundamental than Planck units". The typical length-scale, however, is about twenty orders of magnitude larger than the Planck length. If one were to use this system of constants in the dimensional argument, one would have concluded that quantum gravity already happens at sub-atomic scales---a conclusion that is  empirically incorrect. Wilczek's reason to think these units are more fundamental is that its length unit, unlike the Planck length, does not involve the square root of any of the constants involved. This may not strike everyone as a decisive factor, but then it is unclear anyway how to decide which unit system (if any!) is more fundamental.\footnote{One may also worry that $m_p$ is not a fundamental constant, since protons are not fundamental objects. Wilczek suggests that one can replace $m_p$ with $\Lambda_{QCD}$, a fundamental mass-dimensioned parameter of quantum chromo-dynamics.} Alternatively, one can directly define a length-scale from the cosmological constant, which has dimensions of area---but it is about fifty orders of magnitude smaller than the Planck length! If that is the the true scale of quantum gravity, physicists will find it nearly impossible to obtain empirical data relevant to quantum gravity. It is of course possible that such constructions are ruled out for physical reasons, but that would only emphasise the fact that dimensional analysis alone cannot establish \emph{Planck}.

Faced with these objections, Baez admits that "we cannot prove that the Planck length is significant for quantum gravity." He retreats to the claim that "we can glean some wisdom from pondering the constants $c, G$, and $\hbar$," namely that principles of locality and background independence are essential to any future theory of quantum gravity. But such principles don't follow just from the presence of some constants, and in any case they are unrelated to \emph{Planck}. Indeed, the dimensional argument simply assumes the truth of Inconsistency---the existence of a scale $\ell_{QC}$ at which physics breaks down--- without any justification. I suspect that most physicists ultimately share Baez's belief that dimensional analysis is a heuristic that does not prove much.

%Let me discuss two further responses to these objections. \citet{Rickles2008} concedes in a footnote that "while the dimensional argument cannot demonstrate the necessity of the Planck scale for quantum gravity, it does appear to point to its sufficiency." One way to read this is that the Planck scale is not a boundary but an upper bound, so that quantum gravity may already become important above that scale. But that is clearly not established by dimensional analysis either. Another interpretation is that the occurrence of $G$, $c$ and $\hbar$ in our theories is sufficient but not necessary for the relevance of the Planck scale. The suffiency-claim, however, is merely a restatement of the dimensional argument, so this response would beg the question.\footnote{In private communication, Rickles has suggested that by `the Planck scale' he means a minimum length scale, so that any theory with a minimum length is a theory of quantum gravity but not every theory of quantum gravity has a minimum length. Again, though, that does not establish the quantitative relevance of the Planck length.}

\subsection{Quantum black holes}

Another common case for \emph{Planck} concerns `quantum black holes' \citep{Doplicher1995, Baez2001, Rovelli2007, Rickles2008, Wallace2022b}. The idea is that the Compton wavelength and the Schwarzschild radius coincide at the Planck length. The Compton wavelength, $\ell_C = 2\pi\hbar/mc$, sets a limit on how finely one can localise a particle of mass $m$: at lengths smaller than its Compton wavelength, one will have added sufficient energy to create another particle of the same mass. The Schwarzschild radius, $\ell_S = 2Gm/c^2$, sets another limit on how finely one can localise a particle: at scales smaller than the Schwarzschild radius, a black hole will form behind which the particle remains hidden.

Notice that $\ell_S$ is proportional to $m$, whereas $\ell_C$ scales inversely with $m$. It is possible to increase our localisation by increasing the particle's mass---but only up to a point, for when $m$ becomes too large a black hole will form. These effects cross when $\ell_S = \ell_C$, which is the case for a particle of mass $\sqrt{\pi}m_P$, where $m_P$ is the Planck mass. Both the Compton wavelength and the Schwarzschild radius are then equal to $\sqrt{\pi}\ell_P$ and hence of the order of the Planck length. This result is supposed to support \emph{Planck}.

But there are a number of issues. Firstly, this description of a Planck-scale effect is silent about what happens above the Planck scale. Insofar as quantum black holes indicate the breakdown of our current theories, this breakdown may still occur well above the Planck scale. It can therefore at most establish the Planck scale as a lower bound---but not Boundary. This falls short of \emph{Planck}. Indeed, it is possible that the Planck scale is closer than it appears. In particular, if spacetime has `large' compactified dimensions, then the `fundamental' Planck length---the one obtained from $G$, $c$ and $\hbar$---is much smaller than the `effective' Planck length at which quantum black holes first form. In this scenario, quantum black holes may already form at scales of the order of some TeV \citep{Giddings2002}. This would also mean that the Planck scale is merely a lower bound not a Boundary. The physics here is far from settled, but the very possibility of such scenarios proves that thought experiments of this kind are a fallible heuristic.

Secondly, the immediate significance of quantum black holes seems merely epistemic: they reveal a limit to the precision of our measurements. \citet[1289]{Rovelli2007}, for instance, concludes that "gravity, relativity and quantum theory, taken together, appear to prevent position to be determined more precisely than the Planck scale." But that does not entail that novel physics occurs at the Planck scale. The existence of a limit to our knowledge of the world does not entail Inconsistency, which rather sets a limit to the applicability of our theories. In his PhD thesis, \citet{Wuthrich2006} makes much the same point:

\begin{quote}
Therefore, the argument as given so far must be complemented by a second part asserting that the operationally discrete spacetime at the semi-classical level results from an underlying discreteness at the fundamental Planck level. 
\end{quote}
W\"uthrich is pessimistic about the prospects of such a `second part'. I will discuss this issue further in the next subsection.

Thirdly, even when interpreted not just epistemically the thought experiment does not establish an internal inconsistency. Baez puts the physical relevance of quantum black holes as follows: "the Compton wavelength sets the distance scale at which quantum field theory becomes crucial for understanding the behavior of a particle of a given mass," while "the Schwarzschild radius is roughly the distance scale at which general relativity becomes crucial for understanding the behavior of an object of a given mass" (180). Both theories become important at the Planck scale.

Recall that both external and internal consistency act as constraints on quantum gravity. It seems that here the former notion plays a role: quantum gravity must unify quantum theory and general relativity. But we already have a theory of low-energy quantum gravity that does just that \citep{Wallace2022b}. Therefore, the internal consistency constraint was deemed more relevant for us: current theories are supposed to break down at the Planck scale. The fact that Planck-scale black holes involve both quantum theory and general relativity, however, does not entail Inconsistency.

For a more deflationary perspective on quantum black holes, consider David \citet[215]{Tong2021}'s assessment:

\begin{quote}
General relativity will give you the right answer to any quantum question at energies $E \ll M_P$. But if you throw together two particles at energies $E \gg M_P$, then general relativity will also give you the right answer. That's because, if you throw particles together at very high energies, then you simply form a black hole! [...] This means that, provided we don't do anything stupid, like jump into a black hole, we understand perfectly well what happens in very high energy scattering. You form a big black hole which slowly evaporates over gazillions of years. We never need any knowledge of the fundamental theory of quantum gravity to figure out the physics.
\end{quote}
Our current theories don't fail: they correctly predict that a black hole will form. Of course, we may require new physics to describe what happens inside a black hole or when it evaporates, but that is the case for a macroscopic black hole, too, so it is unrelated to \emph{Planck}.

\subsection{Generalised uncertainty principles}

One of the first physicists to defend the importance of the Planck length was C. Alden Mead \citep{Mead1964}; see \citet{Mead2001} for a brief history. Mead does not directly link the Planck length to a putative theory of quantum gravity. Rather, he aims to show from our current theories that the Planck length is a `fundamental length'. This would mean that those theories contain a fundamental limit, which is closely related to \emph{Planck}.

Mead relies on another thought experiment.\footnote{See \citet{Hossenfelder2013} for a more detailed discussion, and \citet{Adler2010} for further variations on a theme.} The set-up is a modified Heisenberg microscope. Recall that this is an experiment in which we try to measure the position of a massive particle on the $x$-axis by scattering off a photon with frequency $\omega$. The photon is measured by a camera with aperture $\epsilon$. We therefore have:
\begin{equation}
	\Delta x \geq \frac{c}{2\pi \omega \sin \epsilon}.
\end{equation}
We then take into account the gravitational interaction between the photon and the particle. The velocity acquired by the particle is approximately $hG\omega /rc^3$, where $r$ is the effective circumference of the particle, which allows the particle to travel a distance $L = hG\omega/c^4$. Because the direction of travel is unknown, this adds an uncertainty
\begin{equation}
\Delta x \geq \frac{h G\omega \sin \epsilon}{c^4}
\end{equation}
to the particle's position. From (4) and (5), it follows that

\begin{equation}
\Delta x \geq \sqrt{\frac{\hbar G}{c^3}} = \ell_P
\end{equation}
This sort of modification of Heisenberg's uncertainty principle is known as a `Generalised Uncertainty Principle' or `GUP' \citep{Adler1999}. 

Firstly, note that this derivation is clearly not relativistic. This means that $c$ is unbounded, so the uncertainty goes to zero when $c \to \infty$. Furthermore, it does not incorporate general relativity. It is therefore difficult to see how (6) could show that the Planck scale is where quantum \emph{gravity} becomes important. Mead does offer a general-relativistic calculation later in the paper. But that calculation assumes that the uncertainty is bounded from below by the Schwarzschild radius, so it is really a variation on the quantum black hole case discussed above.

Secondly, as Mead himself notes, the result relies on unrealistic assumptions as it violates the conservation of momentum. If one incorporates momentum conservation, the uncertainty increases. Mead notes that this would still entail (6), since the new minimum uncertainty is even greater than the Planck length. The result nevertheless "does not rule out the possibility of some larger fundamental length, but [the Planck length] is to be thought of as a lower limit" (860). Therefore, in the terms of \S3, it at most establishes that quantum gravity has already happened once one reaches the Planck scale, but does not rule out that it may happen well before then. This would mean that we at best have justification for the weak claim that the Planck scale is a lower bound, but not for the stronger claim that it is a Boundary as required by \emph{Planck}.

Finally, like the quantum black holes, the GUP only seems to establish an epistemic limit but not a theoretical one. Mead is clear on this: "whenever the term `fundamental length' is used in this paper, it refers to a length having the physical interpretation discussed here, that is, a limitation on the possibility of measurement" (fn. 10). But as I emphasised in the previous section, a limit to measurement is not necessarily a limit to the applicability of our theories. It does not entail Inconsistency. This would only follow from the kind of operationalism expressed by \citet[928]{Adler2010}: "Since we cannot measure  particle position more accurately than the Planck length, the above result suggests that from an operational perspective the Planck length may represent a minimum physically meaningful distance." That kind of justification would make \emph{Planck} highly controversial indeed.

\citet{Hossenfelder2013} has a less instrumentalist reason to believe that the GUP is not just epistemic: it requires explanation. "Heisenberg's microscope revealed a fundamental limit that is a consequence of the non-commutativity of position and momentum operators in quantum mechanics. The question that the GUP then raises is what modification of quantum mechanics would give rise to the generalized uncertainty" (13). But why should uncertainty principles always demand an explanation? After all, our current physics already seems to fully explain why there are epistemic limits to our measurement: it is just the explanation provided by Mead's calculations. There is a suggestive analogy here, but no substantive justification for \emph{Planck}.

Although the kind of epistemic limit discussed here and in the previous subsection may not establish a breakdown of our theories, it may still seem to suggest that odd stuff happens at the Planck scale. This could constitute weak evidence for new physics at the Planck scale. I don't want to deny all relevance to these thought experiments, but if the above criticisms are correct then their impact is certainly more limited than is often believed.

\subsection{Effective field theories}

I have defined \emph{Planck} as the claim that our current physics breaks down at the Planck scale. We have seen that simple thought experiments cannot establish this. But there is an approach to physics that explicitly concerns the limits of our theories, namely that of \emph{effective field theories}. An effective field theory is only well-defined up to a certain scale, and it is the theory \emph{itself} that tells us up to which scale it is valid. In the words of \citet{Zee2010}: "theories in physics have the ability to announce their own eventual failure and hence their domains of validity." Inconsistency is therefore built in to effective field theories. It is often claimed that quantum gravity, \emph{considered as an effective field theory}, breaks down at the Planck scale \citep[are some examples]{Polchinski1998, Zee2010, Crowther2022}. This would vindicate \emph{Planck} from within the theory itself.

I lack the space to provide a detailed overview of effective field theory, so I will only sketch the main ideas; see \citet{Butterfield2014} or \citet{Williams2022} for introductions aimed at philosophers. For a review of quantum gravity as an effective field theory, see \citet{Burgess2004}.

Let's first consider the idea of \emph{perturbative renormalisation}. In order to obtain a prediction from quantum field theory, we typically expand around a simple vacuum state. However, in many cases these perturbative expansions include terms that sum over all momenta and hence become infinite. Renormalisation is a technique to remove those infinities. In brief, the procedure is to first `regulate' the theory, that is, cut off the integrals at some high but finite momentum. This takes care of the infinities but makes the theory dependent on the cut-off. However, one can then `renormalise' the theory by introducing finitely many `counterterms' in such a way that predictions remain finite when one takes the cut-off to infinity. Unfortunately, this recipe only works for some theories: they are perturbatively renormalisable. If a theory is perturbatively non-renormalisable, one would have to introduce infinitely many counterterms to cancel the infinites.

The general theory of relativity is perturbatively non-renormalisable. Consider the Einstein-Hilbert action:

\begin{equation}
	S_{\text{EH}} = \frac{c}{2\kappa^2} \int d^4x \sqrt{-g} R,
\end{equation}
where $\kappa^2 = \frac{8\pi G}{c^3}$, $g = \det{g_{\mu\nu}}$ and $R$ is the Ricci scalar. We can expand the action around a background field $\eta_{\mu\nu}$:
\begin{equation}
	g_{\mu\nu} = \eta_{\mu\nu} + \frac{\kappa}{\sqrt{c}} h_{\mu\nu}
\end{equation}
This fluctuation represents a graviton. The expansion is complicated, but schematically it looks like:
\begin{equation}
	S = \int d^4 x (\partial h)^2 + \kappa h (\partial h)^2 + \kappa^2 h^2 (\partial h)^2 + ...
\end{equation}
We can see that the expansion increases in powers of $\kappa$.

The second-order terms are non-renormalisable, so in order to cancel the infinities they introduce one would have to add infinitely many counter-terms \citep{Goroff1986}. On the effective field theory approach, this is not a reason to reject the theory. Instead, we infer that general relativity is only valid at length scales that are large compared to $\kappa$. At smaller scales, the theory loses its predictive value. In \citepos[172]{Zee2010} words (note that Zee uses the Planck mass rather than the Planck length):

\begin{quote}
  	Just as in our discussion of the Fermi theory, the nonrenormalizability of quantum gravity tells us that at the Planck energy scale $(1/G_N)^{1/2} \equiv M_{\text{Planck}} \approx 10^{19} m_{\text{proton}}$ new physics must appear. Fermi's theory cried out, and the new physics turned out to be the electroweak theory. Einstein's theory is now crying out. 
  \end{quote}
It would seem that we have derived the Planck length as a theoretical limit. There is a problem, however: $\kappa$ does not have dimensions of length. This is obscured by the widespread use of natural units in particle physics. When $c = \hbar = 1$, both $\kappa = \sqrt{8\pi G}$ and $\ell_P = \sqrt{G}$, so $\kappa = \sqrt{8\pi}\ell_P$, that is, $\kappa$ is of the order of the Planck length. \emph{But this is only the case in this particular system of units}. In other systems of units, such as the standard SI units, the numerical value of $\kappa$ is approximately $7.89 \times 10^{-18} $, which is not at all close to the Planck length. We should not take such unit-dependent relations seriously. Because $\kappa$ is not a length-dimensioned quantity, it cannot determine the length scale at which perturbation theory breaks down.

In order to turn $\kappa$ into a length-dimensioned quantity, one has to multiply it by some constant with the correct dimensions. The square root of Planck's constant has those dimensions. Therefore, $\sqrt{\hbar}\kappa \approx\ell_P$ is a unit-\emph{independent} equation. It would make dimensional sense to use $\kappa$ multiplied by $\sqrt{\hbar}$ as a length-scale. But where does $\hbar$ come from in this equation? The Einstein-Hilbert action only contains $G$ and $c$. There is no reason other than a bit of dimensional analysis: $\hbar$ has exactly the right dimensions. \citet[3]{Polchinski1998} is clear about this:

\begin{quote}
The ratio of the one-graviton correction to the original amplitude must be governed by the dimensionless combination $G_N E^2 \hbar^{-1} c^{-5}$, where $E$ is the characteristic energy of the process; this is the only dimensionless combination that can be formed from the parameters in the problem. [...] From this dimensional analysis one learns immediately that the quantum gravitational correction is an \emph{irrelevant} interaction, meaning that it grows weaker at low energy, and in particular is negligible at particle physics energies of hundreds of GeV. By the same token, the coupling grows stronger at high energy and at $E>M_{Pl}$ perturbation theory breaks down.
\end{quote}
But this is just the `simple dimensional argument' dressed up as a perturbative expansion! 

If the present case is to succeed, we need a physical reason for $\hbar$ to enter the equation. We are interested in \emph{quantum} gravity, so $\hbar$ occurs in the quantisation process. In more detail, on the path-integral approach one defines:

\begin{equation}
	Z = \int \mathcal{D}h_{\mu\nu} e^{iS_{\textit{eff}}/\hbar},
\end{equation}
where $S_{\textit{eff}}$ is an effective action constructed from the Einstein-Hilbert action that includes all terms allowed by the theory's symmetries. Here, $\hbar$ occurs explicitly in the exponential. (Of course, when we couple gravity to matter the exponential must also contain matter terms. I discuss this case below). 

From (10), one can compute physically relevant quantities such as the two-point correlation function:

\begin{equation}
	\langle \Omega | h_{\mu\nu}(x) h_{\alpha\beta}(y) | \Omega \rangle = \int \mathcal{D}h\, h(x) h(y) e^{iS/\hbar},
\end{equation}
which yields the amplitude for a graviton to propagate from $x$ to $y$. It is instructive to see one explicit result based on this approach. From the path integral, one can calculate corrections to the gravitational potential \citep{Donoghue2012}:

\begin{equation}
	V_\text{eff} = -\frac{GMm}{r}\biggl[1 + 3\frac{G(M+m)}{rc^2} + \frac{41}{10\pi}\frac{\hbar G}{r^2c^3} + ... \biggr]	
\end{equation}
The first term is the classical potential; the second term is the correction from general relativity; the third term is a quantum correction, as one can infer from the factor of $\hbar$. This correction, as well as any higher-order corrections, becomes important when $r \leq \sqrt{\hbar G/c^3} = \ell_P$. Unlike in the perturbative expansion of the Einstein-Hilbert action, here $\hbar$ seems to occur naturally as a consequence of the quantisation.

Does this solve the problem, or just push it back one step? After all, we can still ask why we divided $S$ by $\hbar$ in the exponent of $Z$. On the one hand, it seems that the reason is still dimensional: in order to obtain a dimensionless exponent, we must divide $S$ by a constant with dimensions of action---and $\hbar$ is just such a constant. But if that is all that is required, why not divide by $21 \pi \hbar$, or $10^7\hbar$, or by an entirely novel constant with the same dimensions as $\hbar$? Empirically, the difference is unmeasurable: for a distance of 1 fm, the approximate size of a proton, the quantum correction to the potential in (12) is of the order of $10^{-38}$.

On the other hand,  the dimensional analysis is now part of a step within the procedure of path-integral quantisation. This quantisation procedure was used successfully in the development of the standard model. Although we cannot empirically confirm that the same method works for gravity, we can reason on the basis of past success: "given the success quantisation has had as a generator in producing (a) pursuit-worthy, (b) weakly justified and/or (c) empirically confirmed hypotheses in the context of classical theories, quantisation might be expected to produce (A) pursuit-worthy, (B) weakly justified and/or (C) empirically relevant hypotheses [in the context of quantum gravity]" \citep[221]{Linnemann2022}.  In more detail, one explanation for the success of quantisation is that it satisfies the correspondence principle: successor theories should in some sense recover the results of their predecessors. \citet{Saunders1993} phrases it as follows:

\begin{quote}
Innovation proceeds by isolation and independent development of structural features of extant theory. Once entrenched, such heuristics (or canonical forms) are preserved in subsequent developments, and previous theory reformulated in their terms.
\end{quote}
Quantisation is a canonical form in this sense, so we are justified to quantise gravity in the same way as we have successfully quantised other theories.

What does it mean to quantise `in the same way'? Does that just require the same qualitative mathematical framework, or does it also involve the same quantitative values for the constants that appear therein? For the Planck length to appear in the quantisation of quantum gravity, the latter is necessary. This means that $\hbar$ is the relevant constant of action for quantum gravity just as it is for other quantum fields. Such an assumption follows from a broader principle:
\begin{quote}
	\emph{Action-Universality}: every quantum field has the same constant of action, namely $\hbar$.
\end{quote}
We have thereby uncovered Action-Universality as a hidden, as-of-yet unjustified assumption in the claim that quantum gravity happens at the Planck scale. 

The situation becomes more complicated when we consider the coupling of gravity to matter.\footnote{I thank David Wallace for suggesting that I address this case explicitly.} I previously suggested that one could easily have used a different action-dimensioned constant to divide the Einstein-Hilbert action by, but that is not quite true for the full theory. Let $S_M$ and $S_G$ denote the matter and gravitational action respectively. We know that the matter term is divided by $\hbar$: Planck's constant \emph{is} the unit of action for QFT after all. Suppose now that we divide $S_{G}$ by a different constant $k\hbar$, where $k$ is dimensionless. The exponent of the matter+gravity path then looks like:

\begin{equation}
	\frac{S_M}{\hbar} + \frac{S_G}{k\hbar}
\end{equation}
But this is equivalent to

\begin{equation}
	\frac{1}{k\hbar}\Big( kS_M + S_G\Big)
\end{equation}
so that if one were to expand the action in this new constant $k\hbar$, that would amount to an (observable!) multiplication of the matter field by $k$. Put differently, the use of a different constant of action for $S_M$ and $S_G$ affects the dimensionless ratio $S_M/S_G$. This makes it seem as if $G$ has a different value, which is clearly an observable matter.

If one applies the same quantisation procedure to gravity in the presence of matter, then, Action-Universality is forced by empirical considerations. On the one hand, one could take this as evidence for the universality of $\hbar$. It is not the case anymore that one can simply substitute a different constant for $\hbar$; one would have to alter the entire quantisation process. In this case, \emph{Planck} is true \emph{not} because gravity itself is non-renormalisable, as is often claimed, but because it is non-renormalisable \emph{when coupled to matter}. On the other hand, one could think of Action-Universality as a further assumption that underpins the quantisation procedure. It is exactly the assumption of Action-Universality that allows one to take the sum $S_M$ and $S_G$ in the path integral. If we don't have independent reason to believe that Action-Universality is true, then we are not justified in the belief that quantisation will work equally well here as before either. I have no problem with methodological conservatism, but it is also important to know the implicit commitments of a certain formalism or procedure. It turns out that Action-Universality is one of them.

The assessment of Action-Universality itself would require an account of the nature of fundamental constants. For example, one could think of $\hbar$ as like $c$, which is a fundamental feature of the spacetime arena in which physical fields evolve. We would therefore expect any relativistic theory to feature the same constant. (Even this is not so clear, as bimetric theories of gravity feature multiple metrics and hence multiple local upper bounds on two-way speeds. The possibility of such theories is one reason to reject Action-Universality.) But one could equally think of $\hbar$ as more like $G$, which is the coupling constant of a particular force. Different forces have different coupling constants. If $\hbar$ is like this, then it is conceivable that gravity has a different constant of action than other forces, in which case Action-Universality is false. For as far as I am aware, there is no extant account of the metaphysical nature of a constant such as $\hbar$ that could help us decide this question.\footnote{\citet{Jacobs2022} presents an account of $G$ that he believes could also apply to $\hbar$. On this account, a constant is part of a theory's kinematical structure. But this account is neutral as to whether theories can have more than one constant of action.}

We therefore find that the best case for \emph{Planck} relies on a metaphysical principle about the universality of action. Some may endorse Action-Universality on a methodologic basis---we should preserve the quantisation procedure across theories---while others (like me) prefer to keep an open mind, but in any case it is clear that the breakdown of our theories at the Planck scale is not simply a result that is derived from those theories' equations.

\subsection{Beyond the Standard Model}

I have characterised \emph{Planck} as the claim that our \emph{current} theories become internally inconsistent at the Planck scale. We have seen that the best, albeit shaky, case for \emph{Planck} comes from effective field theory. But instead of focus on current theories, one could also try to derive the Planck length from proposed \emph{future} theories. In particular, it is sometimes claimed that the Planck length naturally emerges from `beyond the standard model' theories such as string theory or loop quantum gravity. This claim is strictly irrelevant to \emph{Planck} as construed, since \emph{Planck} concerns the breakdown of low-energy quantum gravity rather than the details of high-energy quantum gravity. Despite this I will briefly discuss why this approach also fails.

Whether beyond-the-standard-model theories posit a fundamental length scale is a question well beyond the scope of this paper. Suppose that it does: is such a fundamental scale then necessarily equal to the Planck scale? This is not the case: one has to fix the fundamental scale `by hand' or on the basis of (currently unavailable) empirical evidence. In her overview of this topic, \citet[66]{Hossenfelder2013} makes this explicit:

\begin{quote}
We have also seen that the minimal length scale is not necessarily the Planck length. In string theory, it is naturally the string scale that comes into play, or a product of the string coupling and the string scale if one takes into account D-branes. Also in [asymptotically safe gravity], or emergent gravity scenarios, the Planck mass might just appear as a coupling constant in some effective limit, while fundamentally some other constant is relevant. We usually talk about the Planck mass because we know of no higher energy scale that is relevant to the physics we know, so it is the obvious candidate, but not necessarily the right one.	
\end{quote}
Let's consider string theory first: its fundamental length-dimensioned quantity, the string length, is "a free parameter that has to be constrained by experiment" (23) and "may differ from the Planck scale" (45). In more detail, the string length is $\ell_S = \sqrt{\alpha'}$, where $\alpha'$ is the so-called Regge slope. The relation between $\ell_S$ and the Planck length, $\ell_P$, is then determined by the dimensionless string coupling constant $g$. The relation is dimension-dependent: in 10 spatial dimensions, $\ell_P = g^{1/4}\ell_S$ \citep{Zwiebach2009}. The relation between the string length and the Planck length is not fixed but depends on the value of $g$. As I briefly noted in \S4.2, it is therefore possible that string theory becomes relevant---contrary to Boundary---well before the Planck scale! Moreover, in some versions of string theory, $g$ is determined by the value of the so-called \emph{dilaton field}. This means that $g$ is dynamical, so it may change both over time and across solutions.

Hossenfelder cautions the same about other proposals, such as non-commutative geometry and causal sets. On the former, she says: "One expects the non-zero entries of [the deformation parameter] to be on the order of about the square of the Planck length, though strictly speaking they are free parameters that have to be constrained by experiment" (41). And on the latter: "This sprinkling has a finite density, which is in principle a parameter, but is usually assumed to be on the order of the Planckian density" (44). In the same vein, \citet{Rovelli1995}, proponents of loop quantum gravity, write that "it is important to note that the Planck constant $l_P$ appearing in [loop quantum gravity] is a bare quantity, that may very well suffer finite renormalizations and thus not coincide with the macroscopic value of $\sqrt{\hbar G/c^3}$" (617). It is plausible that the same is true for other theories: beyond the standard model physics does not support Boundary.

Therefore, as far as I am aware there is no way to directly derive the Planck scale from theories beyond the standard model without input from experiment. The reason to expect that those theories contain Planck-scale parameters, if any, is that our current theories supposedly break down at that scale. But that it just what \emph{Planck} claims, so those theories offer no independent justification for \emph{Planck}.

\section{Conclusion}

I have analysed several arguments for \emph{Planck} and found none of them convincing. The argument from the nonrenormalisability of quantum gravity is the least problematic, but it too relies on a hidden assumption, namely Action-Universality. Therefore, the breakdown of our theories at the Planck scale is not a straightforward prediction of those theories, despite many claims to the contrary.

Although each argument by itself is unsuccessful, one could note that many different arguments point to the same conclusion. The fact that the Planck scale keeps showing up---that it is \emph{robust}---might in itself constitute evidence for \emph{Planck}.  Of course, this type of reasoning does not always work: many bad arguments don't necessarily make a good one. Does it work here? \citet{Linnemann2020} analyses robustness in terms of two further features: concordance and generative entrenchment. Firstly, a feature X of a theory is concordant if and only if "X remains invariant under partially independent multiple determinants" \citep[3]{Soler2012}. But it is questionable whether the arguments for \emph{Planck} are truly independent from each other. For example, dimensional reasoning plays a role in many discussions of the Planck scale, and the possibility of quantum black holes is also used in some derivations of the GUP. Secondly, a feature X is generatively entrenched if and only if "X is involved in an essential way (X plays a quasi-foundational role) in the generation of a huge number of ingredients constituting scientific practices" \citep[3]{Soler2012}. But this is not the case either: while many physicists believe \emph{Planck} to \emph{follow} from our theories, \emph{Planck} seems relatively dispensible as a scientific hypothesis. Although it is a widely-held belief, it is currently not at the centre but at the periphery of the web of beliefs that constitute scientific practice.

Finally, perhaps the Planck scale is merely intended as a heuristic. We simply don't know when quantum gravity happens, but we can entertain some informed speculation. Viewed from this perspective, pathologies such as quantum black holes or the GUP don't indicate the breakdown of our theories, but are indicative of the need for new theories to describe what happens at those scales. This does not mean that such claims are beyond criticism. I concur with \citet[12]{Linnemann2018}, who write in a related context:
\begin{quote}
Let us anticipate another possible objection, namely that it is unfair to appraise arguments which are only meant as mere heuristics or intuitions in the first place. However, we take this point to be ill-founded. That arguments can only be plausibility arguments at the heuristic level does not mean that they are immune to scrutiny and critical assessment tout court. The philosopher of physics' job in the process of discovery of quantum gravity---so we believe---should amount to providing exactly this kind of assessments.
\end{quote}
It is exactly this kind of critical assessment that I hope to have offered. If I am correct then \emph{Planck} is not an established scientific claim, but a tentative hypothesis based on fallible---and perhaps questionable---heuristics.

\section*{Acknowledgements}
I would like to thank Sebastian De Haro, Mahmoud Jalloh, Nat Levine, Tushar Menon, Dean Rickles, David Wallace and two anonymous referees their for feedback and encouragement.

\newpage
\begin{small} %%Makes bib small text size
\singlespacing %%Makes single spaced
\bibliographystyle{psalike} %%bib style found in bst folder, in bibtex folder, in texmf folder.

\thispagestyle{empty} %%Removes page numbers
\end{small} %%End makes bib small text size

\end{document}